\documentclass[twocolumn,superscriptaddress,showpacs,preprintnumbers,amsmath,amssymb,prb]{revtex4}
\usepackage{graphicx}
\usepackage{dcolumn}
\usepackage{bm}
\def\up{|\hspace{-1.0mm} \uparrow \uparrow\rangle}
\def\down{|\hspace{-1.0mm} \downarrow \downarrow\rangle}
\def\updown{|\hspace{-1.0mm} \uparrow \downarrow\rangle}
\def\downup{|\hspace{-1.0mm} \downarrow \uparrow\rangle}

\begin{document}
\title{Proposal of a  full Bell state analyzer for spin qubits in a double quantum dot}

\author{Nobuhiko Yokoshi}
\affiliation{JST-CREST, 4-1-8 Honcho, Saitama 332-0012, Japan}
\affiliation{Nanotechnology Research Institute, AIST,
1-1-1 Umezono, Tsukuba 305-8568, Japan} 
\author{Hiroshi Imamura}
\email{h-imamura@aist.go.jp}
\affiliation{Nanotechnology Research Institute, AIST,
1-1-1 Umezono, Tsukuba 305-8568, Japan} 
\affiliation{JST-CREST, 4-1-8 Honcho, Saitama 332-0012, Japan}
\author{Hideo Kosaka} 
\affiliation{Laboratory for Nanoelectronics and Spintronics, RIEC,
Tohoku Univ., Sendai 980-8577, Japan} 
\affiliation{JST-CREST, 4-1-8 Honcho, Saitama 332-0012,
Japan}

\date{\today}

\begin{abstract}
We propose a scheme for full Bell state measurement of spin qubits in a double quantum dot. Our scheme consists of Pauli spin blockade measurements and biaxial electron spin resonance.  In order to eliminate the average of the Zeeman fields, the double quantum dot is designed so that the Land{\'e} ${\rm g}$-factors of first and second dots satisfy ${\rm g}_1=-{\rm g}_2$ with the use of ${\rm g}$-factor engineering. Thus, we can swap one of the three spin-triplet states for the spin-singlet state without disturbing the other states. Our study shows that the sequential spin-to-charge conversions enable us to implement the full Bell state measurement of electron-spin qubits.
\end{abstract}
\maketitle

Bell state measurement is a key element of quantum information science and technology. It generates an entanglement state of qubits, and plays an important role in quantum teleportation~\cite{Bennett}, entanglement swapping~\cite{Briegel} and quantum key distribution~\cite{Bennett2}. On the other hand, much effort has been devoted to realize solid-state qubits, which are promising candidates for quantum memories and quantum gates. Bell state analyzers for these qubits are then highly desired to explore the potentiality of solid-based quantum devices.

In 2005, Engel and Loss proposed a partial Bell state analyzer for electron spin qubits that determines the parity of two qubits~\cite{Engel}. However the partial analyzer requires some initial source of entanglement for the full Bell state measurement~\cite{Engel}. In this Rapid Communication, we propose a protocol and physical implementation for full Bell state measurements of spin qubits in coupled quantum dots (QDs). Our scheme consists of Pauli spin-blockade measurements~\cite{Ono,Petta,Koppens,Nowack,Laird,Tarucha} and biaxial electron-spin resonance (ESR)~\cite{Koppens,Nowack,Laird,Tarucha,Golovach,Tokura,Rashba}. We show that the sequential spin-to-charge conversions and total spin rotations enable us to implement the full Bell state measurement if the Land{\'e} g-factors of the first and the second QDs are designed to satisfy  ${\rm g}_1=-{\rm g}_2$ with the use of ${\rm g}$-factor engineering. 
\begin{figure}[b]
\includegraphics[width=85mm]{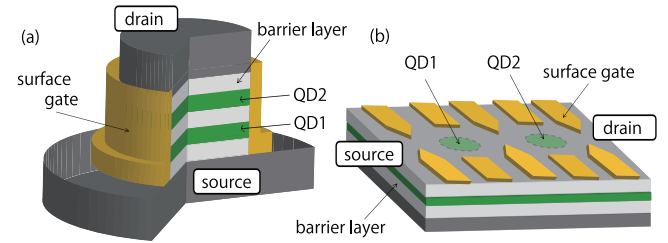} \caption{(color online) Schematic view of (a) vertically and (b) laterally coupled double QDs. Both the QDs are located between the heterostructure barriers, and are then focused to small areas with the use of the surface metallic gates. We assume that the g-factors in the two QDs are designed to satisfy the condition ${\rm g}_1=-{\rm g}_2$. For the vertical double QD, such a condition can be achieved, e.g., by adjusting the thickness and the alloy composition of each quantum well. For the lateral QDs, we can achieve the condition by controlling the equilibrium electron positions in individual QDs when the top and bottom barrier layers have different ${\rm g}$-factors. We take the $z$-axis as the direction of the static magnetic field, and apply oscillating magnetic fields in the $x$- and $y$-directions common to the two QDs. The ac fields flip the electron spins confined in the QDs when the ESR condition is satisfied.}
\label{fig:model}
\end{figure}

The system we have in mind is vertically or laterally coupled QDs as shown in Figs.~\ref{fig:model}(a) and ~\ref{fig:model}(b). We employ the so-called Hund-Mulliken model where the orthonormal state is defined as $\Phi_{1(2)}({\bf r})=(\varphi_{1(2)}-g\varphi_{2(1)})/\sqrt{1-2\Gamma g+g^2}$~\cite{Burkard}. Here $\Gamma$ is the overlap integral of the orbitals in first and second QDs $\varphi_{\rm 1,2}$, and $g=(1-\sqrt{1-\Gamma^2})/\Gamma$. The relevant two-electron states are the $(1,1)$ charge states, $\{|S\rangle, |T_0\rangle, |T_+\rangle, |T_-\rangle \}$, and the spin-singlet $(0,2)$ charge state, $|(0,2)S\rangle$. The label $(m,n)$ refers to the electron number confined in the first and the second QDs, and the Bell basis set is composed of the four (1,1) states, i.e., one spin-singlet state, $|S\rangle=[\updown-\downup]/\sqrt{2}$, and three spin-triplet states, $|T_0\rangle=[\updown+\downup]/\sqrt{2}$ and $|T_{\pm}\rangle=[\up \pm \down]/\sqrt{2}$. Throughout the work, we neglect the spin-triplet $(0,2)$ state because it is energetically inaccessible~\cite{Petta}. The Hamiltonian of the double QD is 
\begin{eqnarray}
\hat{H}_{0}
 \simeq 
 -\epsilon|(0,2)S\rangle\langle(0,2)S|
 +
\sqrt{2}T
\left[
 |(0,2)S\rangle\langle S|
+{\rm H.c.}
\right].
\end{eqnarray} 
Here $\epsilon$ and $T$ represent the gate-controlled detuning energy and the hopping integral both of which are functions of $\Gamma$, and H.c. stands for Hermitian conjugate.

When a magnetic field ${\bf B}$ is applied to the double QD, the Zeeman energy is described by the Hamiltonian $\hat{H}_{{\rm Z}} =\sum_{j=1,2}{\bf h}_{j}\cdot\hat{{\bf s}}_{j}/\hbar$, where $\hat{{\bf s}}_j$ is the spin-$1/2$ operator for the $j$-th QD and ${\bf h}_j=-{\rm g}_j \mu_{\rm B} {\bf B}$ with $\mu_{\rm B}$ being the Bohr magneton. Here the Land\'e ${\rm g}$-factor is defined by ${\rm g}_j=\langle \Phi_j|{\rm g}({\bf r})|\Phi_j \rangle$ since we treat spatially varying ${\rm g}$-factor below. Let us introduce the two-spin operators $\hat{{\bf s}}\equiv \hat{{\bf s}}_{1}+\hat{{\bf s}}_{2}$ and $\delta\hat{{\bf s}}\equiv \hat{{\bf s}}_{1}-\hat{{\bf s}}_{2}$, and the average (inhomogeneity) of the Zeeman fields ${\bf h}=\left({\bf h}_{1}+{\bf h}_{2}\right)/2$ ($\delta{\bf h}=\left({\bf h}_{1}-{\bf h}_{2}\right)/2$). In terms of the Bell basis, $(|S\rangle, |T_0\rangle, |T_+\rangle, |T_-\rangle)$, the Hamiltonian of the Zeeman energy is expressed as
\begin{eqnarray}
\hat{H}_{\rm Z}
= 
\begin{pmatrix}
0 & \delta h^{z} & -i\delta h^{y} & -\delta h^{x}\\ \delta h^{z} & 0 &
h^{x} & ih^{y} \\ i\delta h^{y} & h^{x} & 0 & h^{z} \\ -\delta h^{x} &
-ih^{y} & h^{z} & 0
\end{pmatrix}.
\label{matrixH}
\end{eqnarray}
When the average of the Zeeman field is zero, i.e., ${\bf h}=0$, one can see that Eq.\eqref{matrixH} reduces to
\begin{eqnarray}
\hat{H}_{\rm Z}
=
\delta h^{z}|S\rangle\langle T_{0}|
-i\delta h^{y}|S\rangle\langle T_{+}|
-\delta h^{x}|S\rangle\langle T_{-}|+{\rm H.c.}
\label{hz2}.
\end{eqnarray}
Thus we can sequentially swap one of the triplet states for the $|S\rangle$ state without disturbing the other states by controlling the inhomogeneity $\delta{\bf h}$.

We assume that the system is designed so that the condition ${\rm g}_1=-{\rm g}_2$ is satisfied using g-factor engineering to eliminate the average Zeeman field. In general, an electron ${\rm g}$-factor is a function of band gap and spin-orbit interaction in a semiconductor~\cite{Roth}. Therefore, it is possible to adjust the electron ${\rm g}$-factors by changing the alloy compositions~\cite{Ivchenko,Kosaka3} or well thickness to nearly zero~\cite{Kosaka,Kosaka2}. In addition, the structurally-defined ${\rm g}$-factor is affected also by the penetration of the wave function into the barrier layers~\cite{Jiang}. Then it is further controlled by shifting the equilibrium electron position between the two layers with different ${\rm g}$-factors. Indeed, the ${\rm g}$-factor control~\cite{Jiang} and fine tuning across zero~\cite{Salls} have been achieved using a heterostructure or a quantum well. The ${\rm g}$-factor engineering is applicable also to single electron in a QD. We have actually fabricated the g-factor engineered QDs, and electrically confirmed that the g-factor is as small as designed (${\rm g} = 0.05$)~\cite{Kutsuwa}.

Transition between up and down spins of a localized electron is accomplished using ESR~\cite{Slichter}. Here we take the $z$-axis in the direction of the static magnetic field, i.e., ${\bf B}_0=(0,0,B_0)$. Thus the ac magnetic field ${\bf B}_{\rm ac}(t)={\bf B}_1 \cos \omega t$ in the $x$-$y$ plane makes the electron spin flip when it is resonant with the spin precession. However, it may be challenging to produce strong and oscillating fields in more than one direction. There is another experimental technique, called electric-dipole spin resonance, that uses the electric gate~\cite{Nowack,Laird,Tarucha,Golovach,Tokura,Rashba}. The effective ac field is generated by the cooperation of the oscillating electric field ${\bf E}_{\rm ac}(t)={\bf E}_1 \cos \omega t$ and spin-orbit interaction in the two-dimensional (2D) QDs~\cite{Nowack,Golovach,Rashba}. Within the linear momentum regime, the effective field becomes~\cite{Golovach,Rashba}
\begin{eqnarray}
{\bf B}_{\rm eff}(t)=B_0\frac{l_{\rm d}^2}{\hbar \omega_{\rm d}}  \Bigl[ \frac{e E_{\rm ac}^{x'}(t)}{\lambda_{+}},- \frac{e E_{\rm ac}^{y'}(t)}{\lambda_{-}},0
\Bigr], \label{EDSR}
\end{eqnarray}
where $\hbar \omega_{\rm d}$ $(l_{\rm d})$ is the confinement energy (lateral size) of the 2D QDs, and $x'=(x+y)/\sqrt{2},~y'=(x-y)/\sqrt{2}$. The spin-orbit length is defined as $\lambda_{\pm}=\hbar/m_{\rm e}(\beta\pm\alpha)$, where $m_{\rm e}$ is the electron mass, and $\alpha$ $(\beta)$ is Rashba (Dresselhaus) spin-orbit coupling constant~\cite{SO}.

\begin{figure}[bt]
\includegraphics[width=80mm]{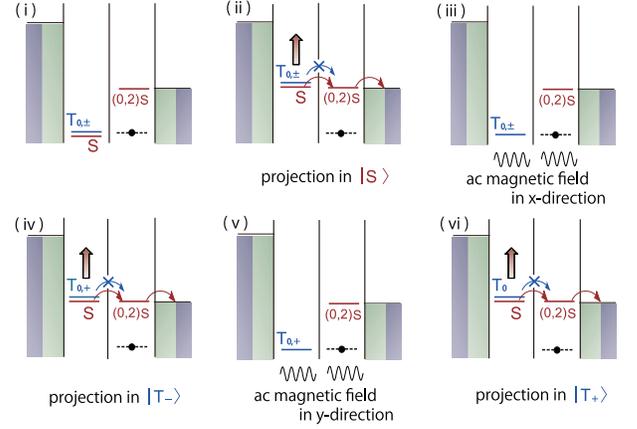} 
 \caption{(color online) The sequence of the Bell state measurement we propose is shown schematically.  The solid lines indicate the energy costs to add an extra electron to the first or the second QD. The dotted line with a filled circle indicates the energy levels of the (0,1) charge state.  (i) Initially the detuning energy is set to be a large negative, $\epsilon=\epsilon_0 \ll -\sqrt{2}T$. (i\hspace{-0.1mm}i) Applying gate voltage to the first QD (depicted in left), we apply the spin blockade regime ($\epsilon=\epsilon_1 >\sqrt{2}T$). The electron in the first QD can tunnel to reservoir when the two-electron state is in $|S\rangle$. (i\hspace{-0.1mm}i\hspace{-0.1mm}i) If no electron tunneling is detected, the system is brought back to the Coulomb blockade regime $\epsilon=\epsilon_0$. We apply the effective ac field $B_{\rm ac}^x(t)$ for the time $\tau_{\rm ESR}$ on both the QDs, and then we wait for the time $t_0$. This swaps $|T_-\rangle$ for $|S\rangle$. (i\hspace{-0.1mm}v) We apply the spin blockade regime to the QDs, and determine whether the two-electron state was initially in $|T_-\rangle$. If the electron tunneling is not detected again, (v) we flip $|T_+\rangle$ into $|S\rangle$ by $B_{\rm ac}^y(t)$, and (v\hspace{-0.1mm}i) determine that the system was in $|T_+\rangle$ or $|T_0\rangle$.}
\label{fig:sequence}
\end{figure}

We show below that manipulating $\epsilon$ and ${\bf B}_{\rm ac}(t)$ enables us to conduct the full Bell state measurement. Initially the detuning energy is set to be large negative, $\epsilon=\epsilon_0 \ll -\sqrt{2}T$, so that the $(0,2)$ charge state lies far above the $(1,1)$ states as shown in Fig.~\ref{fig:sequence}(i). Then, any two-electron state can be expressed by superposing the four Bell states, i.e., $|\psi\rangle=c_s |S\rangle+c_0|T_0\rangle+c_+|T_+\rangle+c_-|T_-\rangle$. For Bell state measurement, it is necessary to perform projection measurements of $|\psi\rangle$ to all the Bell states.

We begin with the projection measurement to $|S\rangle$. We adjust gate voltages on the QDs to apply the spin blockade regime, i.e., $\epsilon=\epsilon_1>\sqrt{2}T$ as in Fig.~\ref{fig:sequence}(i\hspace{-0.1mm}i). When the gate voltage is swept adiabatically for the interdot tunneling $\sqrt{2}T$, the singlet component of $|\psi\rangle$ segues into $|(0,2)S\rangle$ through an avoided crossing. Note that the gate sweep should be nonadiabatic for the inhomogeneity of the Zeeman field $\delta h_0^z=-{\rm g}_1\mu_{\rm B}B_0$ so that the mixing of $|S\rangle$ and $|T_0\rangle$ can be inhibited~\cite{Petta2,Taylor}. Setting the Fermi level of the drain to the energy of $|(0,2)S\rangle$, the interdot tunneling is detected as a current. Because the triplet $(0,2)$ state is neglected, the detection of the electron tunneling determines that $|S\rangle$ is postselected~\cite{Ono,Petta,Koppens,Nowack,Laird,Tarucha}.

In the null result case, the postmeasurement state becomes $|\psi\rangle=N[ c_0|T_0\rangle+c_+|T_+\rangle+c_-|T_-\rangle]$ with normalization factor $N=1/(1-|c_s|^2)^{1/2}$. We thus move on to the next stage, i.e., projection measurement to $|T_-\rangle$. To measure $|T_-\rangle$ via the Pauli spin-blockade measurement described above, we should swap $|T_-\rangle$ for $|S\rangle$. Therefore, we apply the oscillating magnetic field $B_{\rm ac}^x(t)$ common to both the QDs after bringing the energy detuning back to $\epsilon=\epsilon_0$ as shown in Fig.~\ref{fig:sequence} (i\hspace{-0.1mm}i\hspace{-0.1mm}i). In the rotating frame of reference, the two-electron spin state is expressed as
\begin{eqnarray}
|\psi(t)^{\rm r}\rangle =\exp \Bigl[ -i\frac{1}{\hbar^2}\sum_{j=1,2} ({\rm
 g}_j\mu_{\rm B}B_0 \hat{s}_j^z)t \Bigr] |\psi(t)\rangle.
\end{eqnarray}
When the resonance condition, $\hbar\omega=|{\rm g}_j|\mu_B B_0$, is satisfied, the electron spins rotate in the rotating frame as
\begin{eqnarray}
\begin{aligned}
|\hspace{-1mm}\uparrow^{\rm r}\rangle_j &\rightarrow {\mathcal R}_j^x(t)
|\hspace{-1mm}\uparrow^{\rm r}\rangle_j=
\cos\frac{\theta_j}{2}|\hspace{-1mm}\uparrow^{\rm r}\rangle_j
+i\sin\frac{\theta_j}{2}|\hspace{-1mm}\downarrow^{\rm r}\rangle_j, \\
|\hspace{-1mm}\downarrow^{\rm r}\rangle_j &\rightarrow {\mathcal
R}_j^x(t) |\hspace{-1mm}\downarrow^{\rm r}\rangle_j=
\cos\frac{\theta_j}{2}|\hspace{-1mm}\downarrow^{\rm r}\rangle_j
+i\sin\frac{\theta_j}{2}|\hspace{-1mm}\uparrow^{\rm r}\rangle_j,
\end{aligned}
\end{eqnarray}
where $\theta_j=\Omega_j t$ is the rotation angle with Rabi frequency $\Omega_j={\rm g}_j \mu_{\rm B} B_{1}^x/2\hbar$. Note that the spins in the two QDs rotate in opposite directions. If the burst time for the ESR $\tau_{\rm ESR}$ satisfies the condition that $\theta_1(\tau_{\rm ESR})= \pi/2$, $|T_-^{\rm r}\rangle$ is transferred into $|S^{\rm r}\rangle$ without disturbing the other two states. In the laboratory frame of reference, however, the coherent oscillation between $|S\rangle$ and $|T_0\rangle$ evolves due to $\delta h_0^z$. Then, after the ESR sequence, we should let the oscillation evolve for the time $t_0$ which is determined by the condition that $\delta h_0^z (t_0+\tau_{\rm ESR})/\hbar=0~({\rm mod~2\pi})$. As a result, the two-electron state is transferred as $|\psi\rangle\rightarrow {\mathcal R}^z(t_0){\mathcal R}^x(\tau_{\rm ESR})|\psi\rangle= N[c_0|T_0\rangle+c_+|T_+\rangle-ic_-|S\rangle]$. We then perform the Pauli spin blockade measurement to determine whether $|T_-\rangle$ is postselected (see Fig.~\ref{fig:sequence}(i\hspace{-0.1mm}v)). When the null result is gained again, we swap $|T_+\rangle$ for $|S\rangle$ by $B_{\rm ac}^{y}(t)$, and determine which of $|T_+\rangle$ and $|T_0\rangle$ is postselected, as shown in Figs.~\ref{fig:sequence}(v) and \ref{fig:sequence}(v\hspace{-0.1mm}i).

In the preceding discussion, we have assumed that ${\bf h}=0$. However, in the real system, there should be a small mismatch of the absolute values of the ${\rm g}$-factors, i.e., $\delta{\rm g}=-({\rm g}_1$+${\rm g}_2)\neq 0$. This produces a small but finite ${\bf h}$ in Eq.~(\ref{matrixH}), which will reduce the fidelity of the swapping operations. When the ESR frequency is set to $\omega=|{\rm g}_0|\mu_{\rm B}B_0/\hbar$ with ${\rm g}_0=({\rm g}_1-{\rm g}_2)/2$, the mismatch of the ESR frequency in each QD is given by $\delta\omega_{j}=-(-1)^j\delta{\rm g}\mu_{\rm B}B_0/2\hbar$. Here we assume that the frequency mismatch $\delta \omega$ and the Rabi frequency in each QD $\Omega_j$ are much smaller than $|{\rm g}_i|\mu_{\rm B} B_0$, and thus the rotating wave approximation is justified. The ESR Hamiltonian of the double QD becomes~\cite{Slichter}
\begin{eqnarray}
\hat{H}_{\rm ESR}=\sum_{j=1,2} \Bigl( \delta \omega_j \hat{s}^z_j - \frac{{\rm
g}_j \mu_{\rm B}}{\hbar} \frac{{\bf B}_1}{2} \cdot \hat{{\bf s}}_j
\Bigr)
\end{eqnarray}
in the rotating frame. For instance, let us estimate the fidelity of the swapping between $|T_-\rangle$ and $|S\rangle$ states. We apply the oscillating field $B_{\rm ac}^{x}(t)$ to both the QDs for $\tau_{\rm ESR}=\pi/4\Omega_0$, where $\Omega_0=\mu_{\rm B} \tilde{B}/4\hbar$ is the average of the two Rabi frequencies with $\tilde{B}= \sqrt{(\delta{\rm g}B_0)^2+ ({\rm g}_0 B_1^{x})^2}$. The resultant fidelity is calculated as
\begin{eqnarray}
|\langle S^{\rm r}|{\mathcal R}^x(\tau_{\rm ESR})|T_-^{\rm r}\rangle|=
|\cos\Theta| \Biggl\{ 1+{\mathcal O}\Bigl[ \frac{(\delta{\rm g}
B_1^{x})^2}{\tilde{B}^2} \Bigr] \Biggr\}, \label{estimation}
\end{eqnarray}
where $\tan\Theta=\delta{\rm g}B_0/{\rm g}_0B_1^x$ and $\delta{\rm g} B_1^x \ll \tilde{B}$. To achieve the fidelity $\gtrsim 99\%$, we have to control the ${\rm g}$-factors so that $\delta {\rm g}/{\rm g}_0 < 0.01$ when the magnetic fields $\{B_0,B_1\}=\{41 {\rm mT},1.9 {\rm mT} \}$ in Ref.~\onlinecite{Koppens} are employed.

In addition to the ${\rm g}$-factor mismatch, there are a variety of sources for measurement errors. For instance, the spin-orbit interaction causes spin flip through electron-phonon interactions~\cite{Tokura,Khaetskii} and total spin rotation in interdot hopping~\cite{Stepanenko,Yokoshi}. However, the most relevant difficulty is hyperfine interaction with host nuclei. The hyperfine Hamiltonian is given by $\hat{H}_{\rm HF}=\Sigma_{j=1,2}{\bf h}_{{\rm n}j}\cdot\hat{\bf s}_j/\hbar$. The root mean square of the field reaches $|\langle {\bf h}_{{\rm n}j} \rangle_{\rm rms}| \sim 10^{-4} {\rm meV}$ in a QD containing unpolarized $N=10^5$ nuclear spins~\cite{Coish,Koppens2,Lambert}. Since the hyperfine field cannot be controlled by the electron ${\rm g}$-factors, the fluctuating nuclear spins shorten the spin dephasing time $T_2^*$ and decrease the fidelity of the singlet-triplet swappings. 

\begin{figure}[b]
\includegraphics[width=85mm]{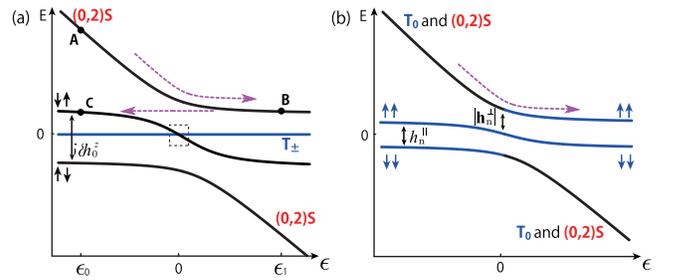} \caption{(color online) (a) Energy diagram for the relevant states. From initial point A with $S(0,2)$, we can reach point C by sweeping the detuning $\epsilon$ adiabatically (A$\rightarrow$B) and then nonadiabatically (B$\rightarrow$C). (b) Enlarged view of the region indicated by square in panel (a). The hyperfine field allows for flip-flop process ($|T_0\rangle \rightarrow \up$) when $h_{\rm n}^{\parallel}>0$. Then driving the detuning up adiabatically from C can reduce the total hyperfine field $h_{\rm n}^{\parallel}$. For $h_{\rm n}^{\parallel}<0$, flip-flop process ($|T_0\rangle \rightarrow \down$) can reduce $|h_{\rm n}^{\parallel}|$ in a similar way.}
\label{fig:diagram}
\end{figure}
It is possible to prolong $T_2^*$ by preparing the nuclei in an approximate eigenstate of $\hat{H}_{\rm HF}$ before we conduct the Bell state measurement. Here we consider the method for dynamic nuclear polarization~\cite{Petta2,Reilly} in the QDs with the ${\rm g}$-factors ${\rm g}_1=-{\rm g}_2$. Note that it is desired to prepare the nuclei so that ${\bf h}_{\rm n}=0$ because the large average Zeeman field is inimical to our Bell state analyzer. The energy diagram in Fig.~\ref{fig:diagram}(a) shows that there exists a state crossing $E=0$~\cite{Taylor,Lambert} which can be expressed as $|0\rangle=\sin\phi|T_0\rangle -\cos\phi|S(0,2)\rangle$ with $\tan\phi=\delta h_0^z/\sqrt{2}T$ at zero detuning. The hyperfine field becomes relevant around $E=0$. The component parallel to the external field, $h_{\rm n}^{\parallel}$, lifts the degeneracy between $\up$ and $\down$, whereas the perpendicular part ${\bf h}_{\rm n}^{\perp}$ opens the gaps nearby the crossing point (see Fig.~\ref{fig:diagram}(b)). We first prepare the system in $|(0,2)S\rangle$ with detuning $\epsilon=\epsilon_0$, i.e., position A in Fig.~\ref{fig:diagram}(a). By sweeping the detuning to $\epsilon_1$ adiabatically and bringing it back to $\epsilon_0$ nonadiabatically, we apply the two-electron state in point C. Then, by driving the system adiabatically along $|0\rangle$, the flip-flop process between $|T_0\rangle$ and $\up$ can occur when $h_{\rm n}^{\parallel}>0$. This process changes the average hyperfine field along ${\bf B}_0$, i.e., reduces $h_{\rm n}^{\parallel}$. When $h_{\rm n}^{\parallel}<0$, on the other hand, it is $\down$ that $|0\rangle$ surges into. As a result, cyclic repetition of these flip-flop transitions leads to a hyperfine field with small ${\bf h}_{\rm n}$ and large $\delta h_{\rm n}^{\parallel}$. Although this method seems complex, any of the processes has been already achieved in experiments using a double QD with ${\rm g}_1={\rm g}_2$~\cite{Petta2,Reilly}.  

As for the time needed in experiment, the burst time is $27$ ns to rotate a single spin by $\pi$/2 for a QD with $|{\rm g}|=0.4$ and $B_1=1.9$ mT~\cite{Koppens}. In each spin-to-charge conversion it takes $\tau \sim 1$ ns to sweep $\epsilon$ for $\sim 1$ meV, adiabatically for $\sqrt{2}T \sim 10^{-2}$ meV~\cite{Petta,Petta2} but nonadiabatically for $|\delta h_0^z| \sim 10^{-3}$ meV~\cite{Koppens}. Such a condition enables us to inhibit the $S$-$T_0$ rotation due to $|\delta h_0^z|$ and the ${\rm g}$-factor modulation by the gate control~\cite{Taylor}. The usage of vertical QDs with large coupling, e.g., $\sqrt{2}T \sim 0.3$ meV~\cite{Ono} can decrease $\tau$ and then freeze the rotation further. Besides, a resonant tunneling to the drain is estimated to take $1.6$ ns in Ref.~\onlinecite{Ono}. Therefore, if the dephasing time $T_2^*$ reaches $\sim 1$ $\mu$s as in the double QD with ${\rm g}_1={\rm g}_2$~\cite{Reilly}, one can verify our scheme in experiment.

In conclusion, we have proposed a full Bell state analyzer for spin qubits in a double QD that are designed to have ${\rm g}$-factors with opposite signs, i.e., ${\rm g}_1=-{\rm g}_2$. We expect that the proposed analyzer enables various devices, for instance, a quantum repeater that is essential for ultra-long quantum communication~\cite{Briegel}. Indeed quantum state transfers from `message carrying' photon qubits to ``storage" matter qubits, e.g., atom ensembles~\cite{atom1,atom2}, $^{31}$P nuclear spins~\cite{Morton} and electron spins in a semiconductor~\cite{Kosaka,Kosaka2} were demonstrated. Although the transfer to electron spins were confirmed in ensemble measurements in a quantum well~\cite{Kosaka}, the mechanism works for single electron in a ${\rm g}$-factor engineered QD with slight modification~\cite{Vrijan}. Combined with these results, our scheme may contribute to constructing quantum information networks as well as processor units.

The authors would like to thank K. Ono, T. Kutsuwa, M. Kuwahara and T. Inagaki for fruitful discussions.

\end{document}